\documentclass[conference]{IEEEtran}

\def\BibTeX{{\rm B\kern-.05em{\sc i\kern-.025em b}\kern-.08em
    T\kern-.1667em\lower.7ex\hbox{E}\kern-.125emX}}

\usepackage{amsmath,amssymb,amsfonts}
\usepackage[noend]{algorithmic}
\usepackage{textcomp}
\usepackage{xcolor}
\usepackage{comment}
\usepackage{algorithm}

\usepackage{graphicx}
\graphicspath{{images/}}
\DeclareGraphicsExtensions{.pdf,.jpeg,.png}
\begin{document}

\title{On the Impact of Device-Level Techniques on Energy-Efficiency of Neural Network Accelerators}%Enhancing Energy-Efficiency in FPGA- and GPU-based\\Neural Network Accelerators}

\author{\IEEEauthorblockN{Seyed Morteza Nabavinejad}
\IEEEauthorblockA{\textit{School of Computer Science} \\
\textit{Institute for Research in Fundamental Sciences (IPM)}\\
Tehran, Iran \\
nabavinejad@ipm.ir}
\and

\IEEEauthorblockN{Behzad Salami}
\IEEEauthorblockA{\textit{Barcelona Supercomputing Center (BSC)} \\
Barcelona, Spain \\
behzad.salami@bsc.es}
}

\maketitle

\begin{abstract}
Energy-efficiency is a key concern for neural network applications. To alleviate this issue, hardware acceleration using FPGAs or GPUs can provide better energy-efficiency than general-purpose processors. However, further improvement of the energy-efficiency of such accelerators will be extremely beneficial specially to deploy neural network in power-constrained edge computing environments. In this paper, we \emph{experimentally} explore the potential of device-level energy-efficiency techniques (e.g., supply voltage underscaling, frequency scaling, and data quantization) for representative off-the-shelf FPGAs compared to GPUs. 
%Undervolting improves the energy-efficiency of FPGAs up to XX\% compared to XX\% on GPUs.
Frequency scaling in both platforms can improve the power and energy consumption but with performance overhead, e.g., in GPUs it improves the power consumption and $GOPs/J$ by up to 34\% and 28\%, respectively. However, leveraging reduced-precision instructions improves power (up to 13\%), energy (up to 20\%), and performance (up to 7\%) simultaneously, with negligible reduction in accuracy of neural network accuracy. 
\end{abstract}


\section{Introduction}\label{sec:introduction}
Deep Neural Networks (DNNs), especially Convolutional Neural Networks (CNNs), are a very popular subcategory of machine learning with significant success rates in classification problems. They are essential for cutting-edge real-world applications, such as image and video classification, speech recognition, and robotics. DNNs learn a model from a dataset during the training phase and in the classification phase, make predictions about a new dataset that has not been seen previously. The computation and the memory requirements of these models are increasing depending on the size of the data as well as the number of layers and the filters. The massive amount of data movement and computational power required for such  models limit their scalability for enterprise applications specially on power-constrained scenarios like drones and mobile devices. 

Hardware acceleration using Graphics Processing Units (GPUs)~\cite{zhang2018shufflenet}, Field Programmable Gate Arrays (FPGAs))~\cite{sharma2016high}, or Application-Specific Integrated Circuits (ASICs)~\cite{sharma2018bit} is a promising solution to overcome such energy-efficiency drawbacks of DNNs.
Among them, FPGAs are efficient data parallel compute engines with increasing use for data centers~\cite{top500}. They attract great attention with their higher power efficiency than GPUs and better flexibility compared to ASICs. Additionally, recent advances in High-Level Synthesis (HLS) tools make it easier to map applications on FPGAs~\cite{putnam2014reconfigurable}. On the other hand, GPUs with internal Single Instruction Multiple Thread (SIMT) structure are convenient to accelerate these DNN models. Many of the emerging deep learning frameworks, such as Caffe~\cite{jia2014caffe}, Torch~\cite{collobert2011torch7}, and TensorFlow~\cite{abadi2016tensorflow} are specifically optimized to run on GPUs with CUDA programming interface. However, the high power consumption of GPUs raises cooling concerns in data centers. Moreover, ASIC-based accelerators are more power- and energy-efficient than FPGAs and GPUs~\cite{nurvitadhi2016accelerating}. However, further improving the energy-efficiency of FPGA- and GPU-based CNN accelerators can enable their deployment in power-constrained environments, as they provide better flexibility than ASICs. Toward this goal, this study aims to comparatively explore the potential of such off-the-shelf hardware platforms to take advantage of energy-efficiency techniques such as supply voltage underscaling, frequency underscaling, and reduced-precision instructions.%, and temperature effect.

We study the following techniques: \textit{\textbf{(i)}} Undervolting, i.e., the supply voltage level underscaling below the nominal level, reduces the total power consumption of the underlying hardware. Thus, it directly leads to the improved energy-efficiency. \textit{\textbf{(ii)}} Frequency underscaling is used to prevent the undervolting-related errors that may occur due to the increased circuit delay when operating at reduced voltage levels. Although, frequency underscaling (simultaneously with supply voltage reduction) may lead to reduced performance but energy-efficiency improves while it prevents the neural network accuracy loss. %\textit{\textbf{(iii)}} Temperature has a direct impact on the static power consumption as well as the reliability of the underlying hardware.
\textit{\textbf{(iii)}} As a common software-level technique, quantization can decrease the precision level of a numerical value used as a weight. This technique aims to reduce the size of CNNs without a significant loss of classification accuracy \cite{parashar2017scnn}. 

Our experimental study is based on a representative Xilinx ZCU102 Ultrascale+ (FPGA) and Nvidia Tesla P40 (GPU). We also perform our study on two state-of-the-art image classification models in the classification phase, i.e., ResNet50 and Inception. On this setup, we evaluate the three above-mentioned energy-efficiency techniques and in detail discuss the efficiency of such techniques. For the GPU accelerator, we deployed the networks on the GPU and applied different techniques to study their effect. %implemented the accelerator on the GPU
For the FPGA-based accelerator, we adopted the experimental results presented in \cite{salami2020experimental}. %For instance, we observed a large voltage guardband below the nominal for FPGA. Eliminating this large guardband leads to more 2.6X energy-efficiency without any performance or reliability overhead. Further undervolting delivers an additional 43\% energy-efficiency improvement; however, it comes with the cost of accuracy loss. To alleviate this issue, we propose an effective simultaneous frequency underscaling technique. This technique prevents this accuracy loss and reduces the power-efficiency gain from 43\% to 25\%. For GPU, we see that the scaling down the frequency can reduce the energy consumption by up to 21\%. Furthermore, employing reduced-precision instructions (INT8) can save up to 25\% energy consumption and improve the performance by up to 13\%, while having negligible impact on accuracy (accuracy loss around 0.5\%).   %\textcolor{red}{(@Moretza: sth about GPU results here)}
The experimental results indicate the surprising finding that unlike FPGA-based accelerator, in GPU-based accelerator the highest power-efficiency and energy-efficiency cannot be simply achieved by highest and lowest frequency, but with careful exploration of all the frequency level options. The right combination of frequency level and precision of the models can lead to up to 35\% power-efficiency and 28\% energy-efficiency improvement. The temperature of the GPU accelerator can also be reduced by around 20\% using the lowest frequency level, to meet certain limitations in temperature-critical situations in embedded systems. 

To our knowledge, this is the first paper to simultaneously evaluate several energy-efficiency hardware- and software-level techniques on FPGA- and GPU-based neural networks. We make the following major contributions in this paper:
\begin{itemize}
    \item We explore the effect of quantization and employment of reduced-precision instructions on the performance of DNNs as a function of computational complexity of the DNNs. We observe that more complex DNNs benefit from the reduced-precision instructions more than less complex ones in terms of runtime, power, and energy consumption.
    \item Conducting extensive set of experiments, we study the impact of DVFS on runtime, power, energy, and temperature of the networks. We  show that the difference between runtime under lowest and highest GPU frequency is significant. So when the DVFS is reduced, it is possible to employ reduced-precision instructions to improve the runtime and mitigate the negative impact of low frequency.
    \item We evaluate the potential of energy-efficiency techniques on FPGA and GPU architectures while running a neural network applications.
    \item Comparing FPGA and GPU regarding frequency scaling, we find that in FPGA the best results for various metrics can be obtained at highest or lowest level of frequency. However, in GPU, more careful exploration of available frequency levels is needed since the best results for some metrics is achievable at frequency levels other than highest and lowest one.   
\end{itemize}

The rest of the paper is organized as follows: In Section \ref{sec:gpu}, we present and discuss the methodology and experimental results for our GPU-based CNN accelerator. In Section \ref{sec:FPGA-GPU}, we compare and discuss the energy-efficiency results between FPGA- and GPU-based accelerators. In Section \ref{sec:related}, we review the related works and conclude the paper in Section \ref{sec:conclusion}.

\section{GPU-based DNN}\label{sec:gpu}

\subsection{Background}\label{subsec:gpu_back}

Various Machine Learning (ML) frameworks are presented such as Caffe \cite{jia2014caffe}, Torch \cite{collobert2011torch7}, and TensorFlow~\cite{abadi2016tensorflow}. TensorFlow employs dataflow graphs to represent computation pattern of ML approaches including DNNs. It maps the dataflow graph nodes on available computing cores of a single machine or a cluster of machines. Hence, it can support large-scale and heterogeneous clusters or hardware with hundreds of computing cores such as GPUs. It can also work with other hardware platforms including multi-core CPUs and ASICs. While TensorFlow focuses on deep neural networks, it can support a variety of other ML applications as well~\cite{abadi2016tensorflow}.

TensorRT platform is developed and extended on top of CUDA to optimize the inference phase DNNs and consists of an inference optimizer and a runtime library. A trained DNN in the form of a frozen graph with its parameters is the input of TensorRT. TensorRT generates an optimized inference engine based on the frozen graph of DNN. TensorRT leverages a set of methods and techniques such as graph optimization and layer fusion for optimizing the inference engine of DNN. It can also generate reduced-precision versions of the inference engine that are able to employ the reduced-precision instructions of GPU families that provide architectural support for such instructions. Deep learning inference applications such as image classification and video streaming can leverage various reduced-precision versions of DNNs such as FP32, FP16, and INT8 that are offered by TensorRT to improve the inference latency, power, and energy, provided that the GPU supports those precision~\cite{NVIDIATensorRT}.

\subsection{Methodology}\label{sec:gpu_method}

\begin{figure}
\centering
\includegraphics[width=.8\linewidth]{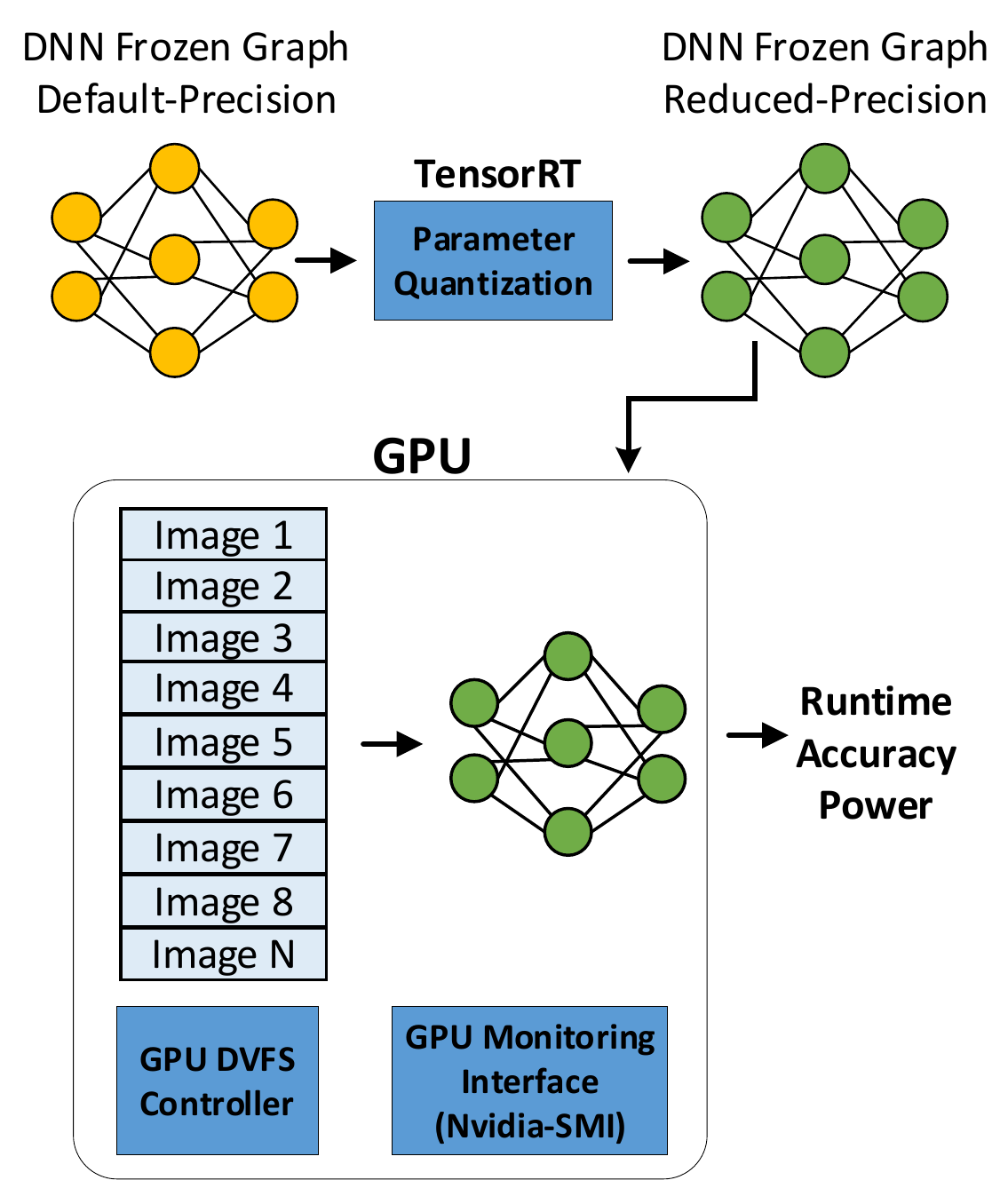}
\caption{Flow of experimental results. First, TensorRT is employed to generate the FP32/INT8 version of the DNNs. Then, they are deployed on GPU for execution. During the execution, the various metrics such as power consumption are monitored using nvidia-smi tool. We use the same tool to apply DVFS on GPU and conduct the related experiments. Finally the labels generated for images by network are compared against the original ones provided by the dataset to evaluate the accuracy of results.}
\label{fig:flow}
\end{figure}

The steps that we follow to conduct the experiments are presented in Fig. \ref{fig:flow}. We use the frozen graph model of DNNs that is generated for inference. The frozen graph is generated from the pre-trained model that is trained with ImageNet dataset \cite{ILSVRC15}. To have FP32 and INT8 precision models of the network \footnote{In our experiments, we also considered 16-bit Floating-Point (FP16)
precision. However, we saw that there is almost no difference between FP32
and FP16 regarding inference time. The reason is that our GPU does not
support half precision arithmetic in its architecture}, we first generate their counterpart frozen graph with TensorRT and then deploy them on the GPU. After that, the frozen graph starts execution on GPU to classify the images and tag them with the predefined labels. The predefined set of labels includes 1000 label classes from ImageNet dataset. During the execution, the desirable metrics such as power consumption are monitored and collected using the GPU Monitoring Interface (NVIDIA System Management Interface- nvidia-smi \cite{nvidiasmi}). Upon completion of inference, the labels generated for each image are compared to original labels provided by dataset to calculate the accuracy of the results. The performance reported for the networks throughout the paper is based on the inference of all the images. The GPU Dynamic Voltage and Frequency Scaling (DVFS) controller is also used to control the frequency of GPU.

\subsubsection{Quantization Technique}

In our work, we use TensorRT to quantize the weights
from FP32 to INT8. TensorRT is developed to optimize
the inference phase of deep learning applications. TensorRT
optimizes neural network models that are trained in major
frameworks such as TensorFlow by calibrating the weights
for lower precision with negligible accuracy loss. Symmetric
linear quantization is used to scale FP32 to INT8. A saturation
threshold for scale factor is considered and any value above
(below) that threshold is mapped to +127 (-127) (max range
of INT8), and the rest of values are mapped to a value
between -127 and +127. To find the best threshold, FP32
inference is run on a calibration dataset. After that, various
quantized distributions with different saturation thresholds are
generated. The threshold that minimizes the information lost
is selected using Kullback-Leibler divergence. Next, FP32
weights are quantized to INT8 using the selected threshold.
Finally, calibration table and INT8 frozen graph are generated
for inference.
For the Inception network, the accuracy for FP32 model is 88.49\% and for INT8 model is 88.24\%. For the ResNet50 network, the accuracy for FP32 and INT8 models is 87.99\% and 87.4\%, respectively. These results are for the GPU implementation of the networks.

\subsection{Evaluation Platform}\label{subsec:gpu_evalPlat}

We run our experiments on a dual-socket Xeon server. It has two  E5-2680 v4 Xeon chips where each of the chips has 28 cores running at 2.4 GHz. The server has 128 GB of DDR4 memory. Ubuntu 16.04 with kernel 4.4 is installed on the server with the python 2.7, CUDA 10.0, TensorFlow 1.15, and TensorRT 4.0. The server is equipped with a PCI Express Gen3 Nvidia Tesla P40 GPU Accelerator. The Tesla P40 leverages Nvidia Pascal architecture and has 3840 CUDA cores. The total memory capacity of GPU is  24 GB GDDR5 memory and its maximum power limit is 250W. The GPU accelerator supports both FP32 and INT8 instructions, optimized for deep learning inference \cite{NVIDIAP40}. The GPU has 79 DVFS levels, starting from 544 MHz to 1531 MHz. Levels are very close to each other and moving from one level to another has negligible effect on GPU performance and power consumption. So, we chose 10 levels that have more distance from each other, and consequently, their effect is more significant: 544 MHz, 632 MHz, 734 MHz, 835 MHz, 949 MHz, 1063 MHz, 1189 MHz, 1303 MHz, 1430 MHz, and 1531 MHz. GPU vendor only allows controlling the frequency, but not the voltage. It adjusts the voltage according to selected frequency via the GPU driver, automatically.\par

\subsection{Experimental Results}\label{subsec:gpu_expres}
\begin{figure}
\centering
\includegraphics[width=.8\linewidth]{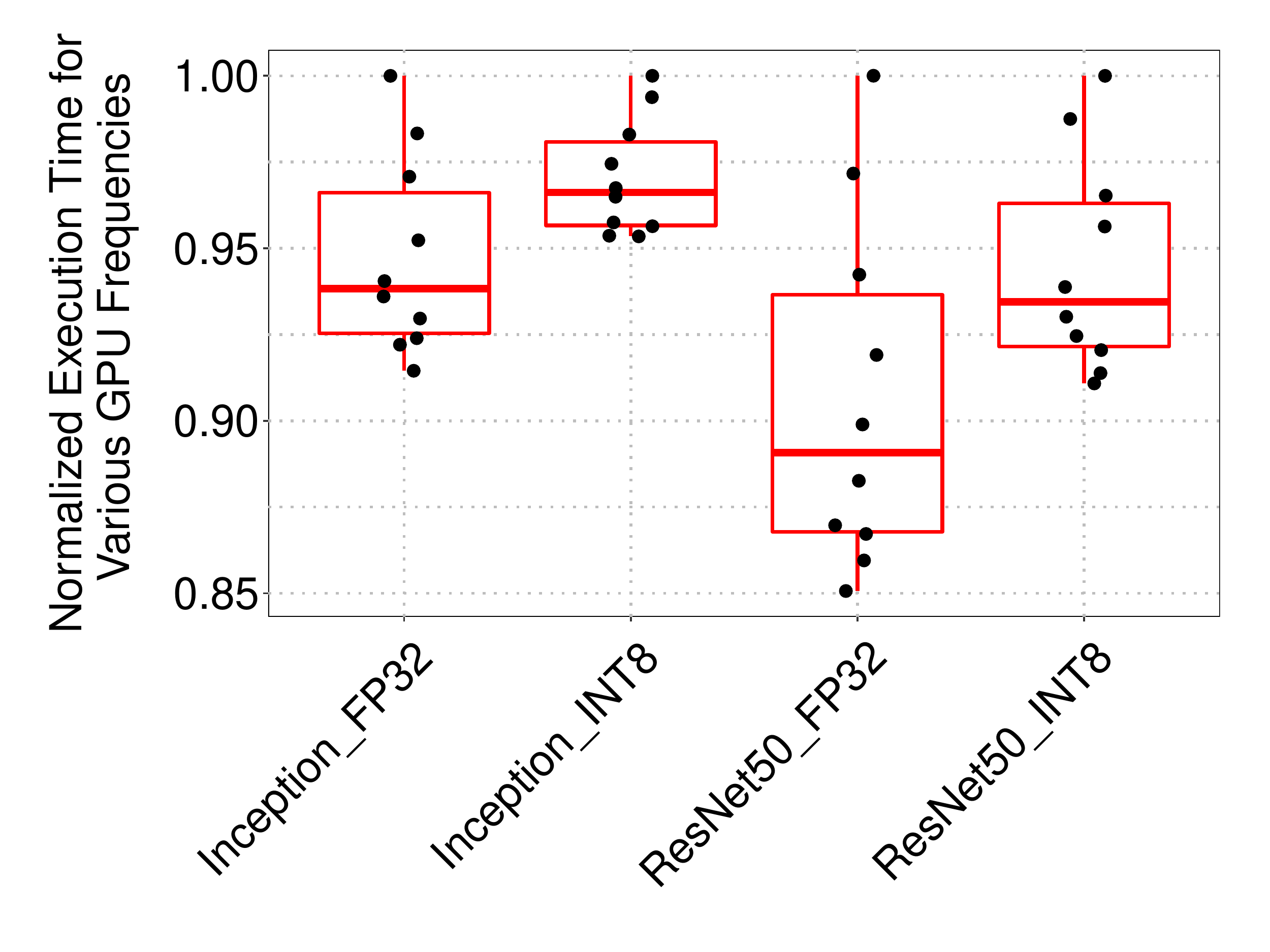}
\caption{Execution time variation of networks under different frequencies for both precision.}
\label{fig:gpuRuntimeVariation}
\end{figure}

In this section, we present the experimental results to study the impact of DVFS on performance, accuracy, and power consumption of networks under various precision. For each network, we consider FP32 and INT8 precision, apply various GPU DVFS levels, and measure the performance, accuracy, and power consumption. As mentioned in previous section, we choose 10 DVFS levels of GPU. The performance results are depicted in Fig. \ref{fig:gpuRuntimeVariation}. The performance of networks for processing 50000 images is measured under different DVFS settings for each precision, and normalized to the lowest performance of that precision (performance of the lowest DVFS level, i.e., 544 Mhz in that specific precision). The black dots indicate the performance for each DVFS level and the horizontal lines of each box show the 25th, 50th, and 75th percentile of dots. For both networks, we see notable performance fluctuation as we change the frequency of GPU. However, the performance variation is more significant in FP32 precision than INT8. We see that INT8-precision can improve the performance that is affected by DVFS. In both networks, the INT8-precision can mitigate the impact of DVFS remarkably.\par

Moreover, we see that the performance of ResNet50 is more sensitive to frequency scaling than Inception in both precision. We can conclude that the effect of DVFS on performance of DNNs has a direct relationship with their architecture. Since DVFS mainly affects the computation time rather than data transfer time, larger and more complex networks are more affected by frequency scaling of GPU. The computation time relative to the overall execution time is higher for larger networks in comparison to small ones, and hence, they are more prone to performance degradation due to DVFS. We conclude that employing reduced-precision instructions can help networks to resist against frequency scaling and reduce its negative impact. It is very useful for systems that have specific Service Level Agreements (SLA) and should reach a certain level of Quality of Service (QoS) in the form of a predefined tail latency or other similar metrics.

\begin{figure}
\centering
\includegraphics[width=\linewidth]{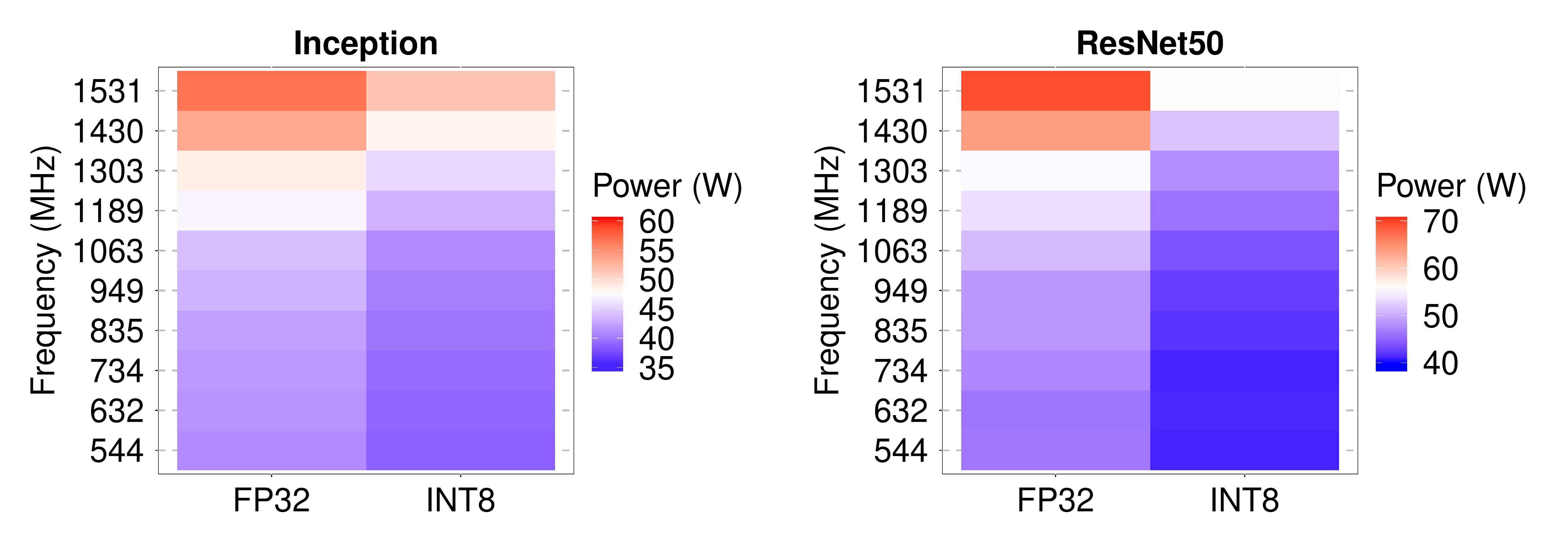}
\caption{Power consumption of GPU for processing the images in each DNN when applying different frequency.}
\label{fig:gpuPower}
\end{figure}

\begin{figure*}
\centering
\includegraphics[width=.9\linewidth]{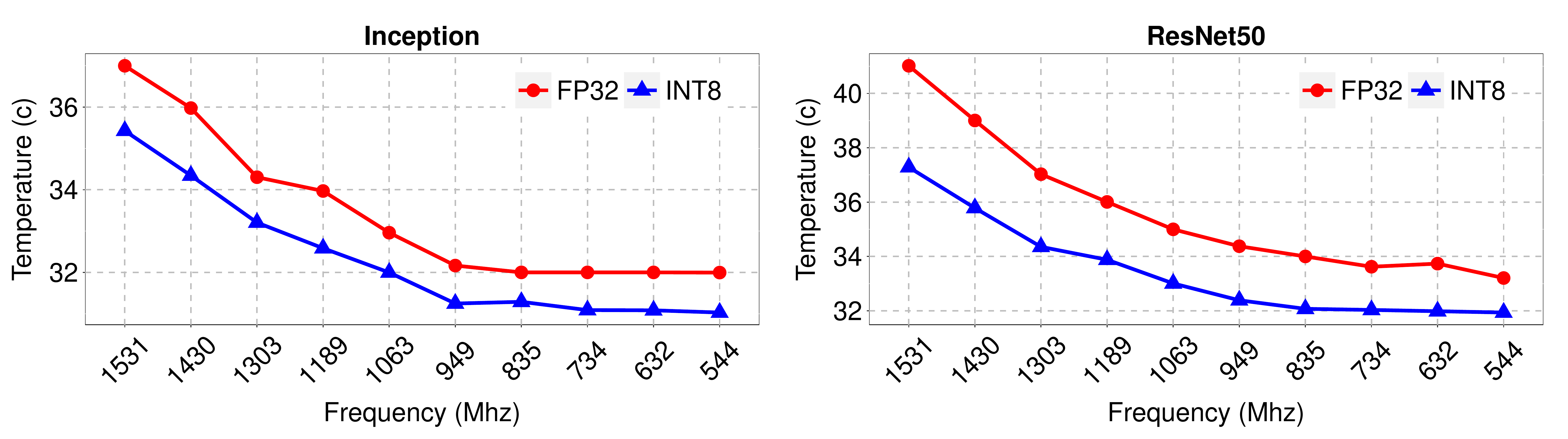}
\caption{Trend of GPU temperature under various frequencies.}
\label{fig:gpuTemperature}
\end{figure*}

Obviously, the power consumption of GPU is also affected by DVFS. The results for power consumption are depicted in Fig. \ref{fig:gpuPower}. As expected, the power consumption decreases with the frequency underscaling. We also observe that the INT8-precision instructions lead to lower power consumption than FP32. Hence, we can conclude that combining low frequency and reduced precision can yield more power reduction than traditional frequency scaling approach. Finally, Fig. \ref{fig:gpuTemperature} shows the effect of frequency scaling, hence the power consumption,  on GPU temperature for both precision. We see that INT8 precision leads to decreased GPU temperature. The temperature has almost the same pattern with power and reduced-precision can help to further decrease it, in addition to frequency underscaling.

Similar to results presented for FPGA in Table 2 of \cite{salami2020experimental}, here we report Giga Operations per second ($GOPs$), power-efficiency ($GOPs/W$), and energy-efficiency ($GOPs/J$) for both Inception and ResNet50 in Table \ref{table:inceptio} and \ref{table:resnet}, respectively. All results are normalized to the highest frequency (1531 Mhz). While the best value of $GOPs$ is obtained with the highest frequency setting, the best value of power consumption is achieved with the lowest frequency setting. However, for $GOPs/W$ and $GOPs/J$, a frequency setting different than the highest and lowest ones has better results based on the network and the precision.

\begin{table}[]
\caption{Evaluation of voltage frequency scaling of GPU for Inception for both precision. Best results for each metric is marked in \textcolor{blue}{blue}.}
\centering
\setlength{\tabcolsep}{3pt}
\label{table:inceptio}
\begin{tabular}{c|cc|cc|cc|cc}
                                                          & \multicolumn{2}{c|}{GOPs} & \multicolumn{2}{c|}{Power (W)} & \multicolumn{2}{c|}{GOPs / W} & \multicolumn{2}{c}{GOPs / J} \\ \hline
\begin{tabular}[c]{@{}c@{}}Frequency\\ (Mhz)\end{tabular} & FP32        & INT8       & FP32           & INT8          & FP32          & INT8          & FP32          & INT8          \\ \hline
1531                                                      & \textcolor{blue}{1}           & \textcolor{blue}{1}          & 1              & 1             & 1             & 1             & 1             & 1             \\
1430                                                      & 0.992       & 0.997      & 0.941          & 0.937         & 1.054         & 1.065         & 1.045         & 1.062         \\
1303                                                      & 0.99        & 0.996      & 0.857          & 0.878         & 1.154         & 1.135         & 1.143         & 1.13          \\
1189                                                      & 0.984       & 1          & 0.826          & 0.838         & 1.191         & 1.193         & 1.172         & 1.193         \\
1063                                                      & 0.977       & 0.988      & 0.772          & 0.794         & 1.265         & 1.244         & 1.236         & 1.23          \\
949                                                       & 0.972       & 0.986      & 0.762          & 0.778         & 1.277         & 1.267         & 1.241         & \textcolor{blue}{1.249}         \\
835                                                       & 0.96        & 0.979      & 0.742          & 0.769         & \textcolor{blue}{1.294}         & 1.272         & \textcolor{blue}{1.243}         & 1.245         \\
734                                                       & 0.942       & 0.97       & 0.734          & 0.757         & 1.284         & \textcolor{blue}{1.282}         & 1.209         & 1.244         \\
632                                                       & 0.93        & 0.96       & 0.728          & 0.752         & 1.277         & 1.277         & 1.188         & 1.225         \\
544                                                       & 0.914       & 0.954      & \textcolor{blue}{0.722}          & \textcolor{blue}{0.748}         & 1.266         & 1.276         & 1.158         & 1.217         \\ \hline
\end{tabular}
\end{table}

\begin{table}[]

\caption{Evaluation of voltage frequency scaling of GPU for ResNet50 for both precision. Best results for each metric is marked in \textcolor{blue}{blue}.}
\centering
\setlength{\tabcolsep}{3pt}
\label{table:resnet}
\begin{tabular}{c|cc|cc|cc|cc}
                                                          & \multicolumn{2}{c|}{GOPs} & \multicolumn{2}{c|}{Power (W)} & \multicolumn{2}{c|}{GOPs / W} & \multicolumn{2}{c}{GOPs / J} \\ \hline
\begin{tabular}[c]{@{}c@{}}Frequency\\ (Mhz)\end{tabular} & FP32        & INT8        & FP32           & INT8          & FP32          & INT8          & FP32          & INT8         \\ \hline
1531                                                      & \textcolor{blue}{1}           & \textcolor{blue}{1}           & 1              & 1             & 1             & 1             & 1             & 1            \\
1430                                                      & 0.99        & 0.997       & 0.915          & 0.921         & 1.081         & 1.082         & 1.07          & 1.078        \\
1303                                                      & 0.981       & 0.99        & 0.796          & 0.852         & 1.232         & 1.161         & 1.209         & 1.149        \\
1189                                                      & 0.978       & 0.985       & 0.769          & 0.817         & 1.272         & 1.205         & 1.245         & 1.188        \\
1063                                                      & 0.964       & 0.979       & 0.734          & 0.778         & 1.314         & 1.259         & 1.266         & 1.233        \\
949                                                       & 0.946       & 0.97        & 0.697          & 0.756         & \textcolor{blue}{1.358}         & 1.284         & \textcolor{blue}{1.285}         & \textcolor{blue}{1.246}        \\
835                                                       & 0.926       & 0.952       & 0.694          & 0.742         & 1.333         & 1.284         & 1.234         & 1.223        \\
734                                                       & 0.903       & 0.944       & 0.68           & 0.727         & 1.328         & \textcolor{blue}{1.297}         & 1.199         & 1.224        \\
632                                                       & 0.875       & 0.922       & 0.665          & 0.73          & 1.318         & 1.263         & 1.154         & 1.165        \\
544                                                       & 0.851       & 0.911       & \textcolor{blue}{0.664}          & \textcolor{blue}{0.727}         & 1.279         & 1.253         & 1.088         & 1.141        \\ \hline
\end{tabular}
\end{table}

\section{FPGA- vs. GPU-based CNN Accelerators} \label{sec:FPGA-GPU}
In this section, we comparatively evaluate the impact of the energy-efficiency techniques for FPGA- and GPU-based CNN workloads, as described in section~\ref{sec:gpu}. Note that for this comparison, we use the results presented in~\cite{salami2020experimental} for FPGA-based DNNs, as we find it a fair comparison case. This is because, \cite{salami2020experimental} is based on modern representative FPGA platforms of Xilinx ZCU102, uses the similar DNN models with this paper, and evaluates the energy-efficiency techniques, which are considered in this paper as well.   

\subsection{Performance Analysis Through Voltage and Frequency Underscaling}
Based on the results presented in Table 2 of \cite{salami2020experimental}, Table \ref{table:inceptio}, and Table \ref{table:resnet}, we compare the impact of voltage and frequency scaling on the performance (i.e., $GOPs$) of FPGA and GPU running CNN workloads. More specifically, we observe that:
\begin{itemize}
    \item in the FPGA case, the highest frequency is nearly 50\% more than the lowest frequency. However, in the GPU case, the highest frequency is almost three times of lowest one. The reason is that in the FPGA implementation of this technique, the minimum supply voltage that we could practically underscale is limited to $V_{crash}$, which FPGA stops operating below it.
    \item for the FPGA case, the $GOPs$ varies by 30\% from highest to lowest frequency. But for the GPU case, the $GOPs$ variation is less significant. The highest variation happens for ResNet50 in FP32 precision, where the $GOPs$ gap is around 15\%, and the lowest gap is 5\% for Inception in INT8 precision. For both networks, we see that $GOPs$ variation in INT8 is less than FP32, which means that we can expect less performance variation in the presence of reduced-precision instructions.
    \end{itemize}

\subsection{Power Consumption Analysis Through Voltage and Frequency Underscaling}
For power consumption, again we see a wider gap in the FPGA (i.e., almost 46\%) than the GPU case (i.e., 34\% highest in ResNet50). Relatively narrower gap in INT8 than FP32 can be seen in GPU results as well. The power gap difference observed between FPGA and GPU can justify the difference between $GOPs$ gap discussed earlier. The wider gap in power consumption can translate to wider computing power gap, which leads to wider $GOPs$ gap. Since the power gap between the lowest and the highest frequency is wider in FPGA than GPU, we see a wider gap between $GOPs$ of the lowest and the highest frequency in FPGA as well.

\subsection{Power- and Energy-efficiency Analysis Through Voltage and Frequency Underscaling}
The first paragraph of this section would talk about the figure and the second would explain it. \par

\begin{figure*}
\centering
\includegraphics[width=.9\linewidth]{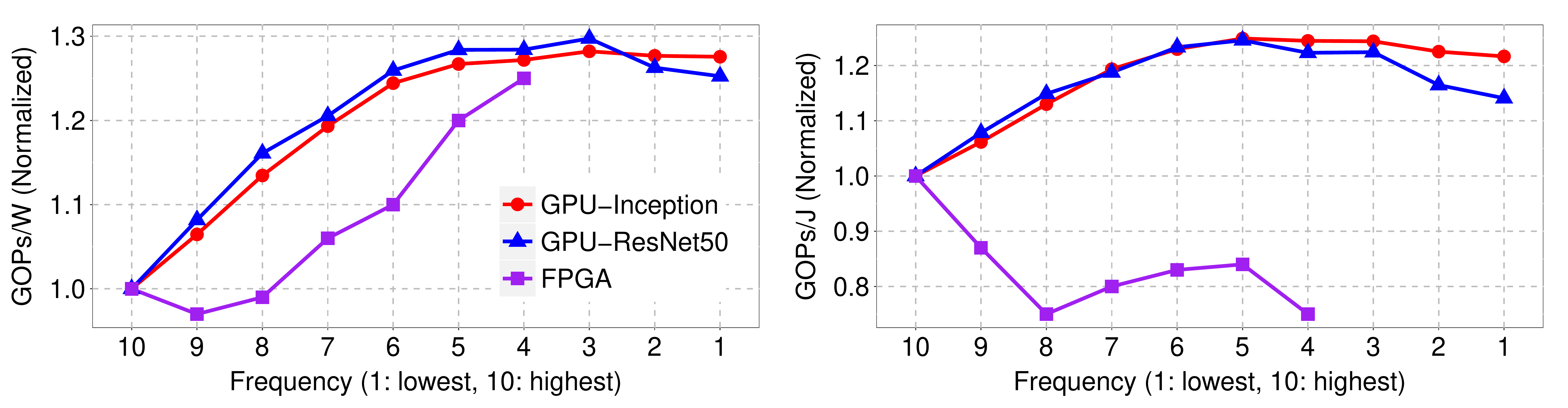}
\caption{Effects of frequency underscaling in power- and energy-efficiency of FPGA- vs. GPU-based CNN accelerator} %(frequency levels are shown normalized).}
\label{fig:gpuvsfpga}
\end{figure*}

Fig. \ref{fig:gpuvsfpga} shows the different behavior of GPU and FPGA regarding $GOPs/W$ and $GOPs/J$. In this figure, we have only depicted the INT8 results for GPU. Since the frequency levels in FPGA (7) is less than GPU (10), only seven points for FPGA are depicted. Finally, since the value of frequency levels in FPGA and GPU are different (for GPU from 544 MHz to 1531 MHz, and for FPGA from 200 MHz to 333 MHz), we have used numbers from 1 to 10 to show frequency levels, instead of the exact values.\par

While the best results for $GOPs$ and power is obtained by the same level of frequency in both FPGA and GPU (the highest frequency for $GOPs$ and the lowest one for power), for $GOPs/W$ and $GOPs/J$ the pattern is different. In the FPGA case, the best result for $GOPs/W$ is obtained by the lowest frequency and for $GOPs/J$ by the highest one. These results indicate that FPGA-based NN designs with frequency scaling has the optimal energy operation point at lowest frequency while the optimal energy-delay operation point is at the highest frequency. In contrast with FPGA, for GPU the frequency that yields the best performance-power and performance-energy ($GOPs/W$, $GOPs/J$) trade-off is between the lowest and highest frequency. We see in Table \ref{table:inceptio} and Table \ref{table:resnet} that the best frequency for $GOPs/W$ and $GOPs/J$ depends on both network and precision. These results indicate the non-linear behavior of GPU regarding frequency scaling. Simply selecting the highest or lowest frequency cannot always guarantee the best results. 

\subsection{Temperature Effects for FPGA- vs GPU-based CNN Accelerator}
In Fig. 10 of \cite{salami2020experimental}, we see that the temperature has a direct impact on the accuracy of FPGA results in critical region. As the temperature increases, the accuracy increases too. However, for GPU, we do not observe any relation between temperature and accuracy. While the temperature increases as the frequency increases, the accuracy of both networks remains constant. The reason is that in the critical region of the FPGA, the voltage is so low that the effect of temperature on latency, and consequently the error rate, of circuit is significant. However for GPU, the voltage is still in guardband region even for very low frequencies, and hence, the temperature does not affect the accuracy of the networks. As mentioned in section~\ref{subsec:gpu_evalPlat}, there is no interface for controlling the voltage of GPU. Therefore, we cannot set the voltage of GPU in critical region to further study the effect of temperature on accuracy.

\section{Related Work} \label{sec:related}
Energy-efficiency is an important concern specially for power-constrained modern applications like DNNs. Energy-efficiency, specially in the hardware-level, is studied either using simulators or on the real underlying hardware. For instance, about undervolting as one of the energy-efficiency techniques evaluated in this paper, majority of the simulation-based studies target CPUs ~\cite{roelke2017pre,yalcin2016exploring,swaminathan2017bravo}, specifically designed for caches~\cite{alameldeen2010adaptive,wilkerson2008trading,wilkerson2010reducing} or branch predictors~\cite{chatzidimitriou2019assessing}. There are also studies for ASIC CNN accelerators~\cite{reagen2016minerva,zhang2018thundervolt}. Additionally, there are FPGA-based designs either fully in  simulation~\cite{mottaghi2019aging} or emulation of FPGA netlists on simulation frameworks~\cite{khaleghi2019fpga,salamat2019workload}. Conducting similar analysis on the real hardware requires a moderate engineering effort regarding physical limitations, but provides relatively more accurate results. For instance, there are several undervolting approaches targeting real hardware system components, such as CPUs~\cite{bacha2013dynamic,papadimitriou2017harnessing}, GPUs~\cite{zou2018voltage, leng2015gpu}, ASICs~\cite{chandramoorthy2019resilient,pandey2019greentpu}, DRAMs~\cite{chang2017understanding, chang2018voltron}, and FPGAs~\cite{salami2018comprehensive, salami2020experimental}. This paper complements the second approach on the real hardware by a more comprehensive experimental study of several state-of-the-art energy-efficiency techniques on GPUS and also, compare it with FPGAs, while running DNN applications. 
 DVFS is a common method to manage the power consumption of a processor via tuning the frequency of clock, and consequently, supply voltage. Controlling the power consumption and performance of GPUs using DVFS has been studied in a large body of research \cite{jiao2015improving,zamani2019greenmm, guerreiro2019dvfs}. Jiao et al. \cite{jiao2015improving} leveraged the capability of modern GPUs for hosting concurrent kernels, along with GPU core and memory frequency scaling to improve the performance per watt. Guerreiro et al. \cite{guerreiro2019dvfs} proposed classification models for impact of DVFS on performance and power of GPU applications. Based on the models, they aimed to predict the impact of different DVFS levels on the applications and set the frequency of GPU accordingly. To reduce the energy consumption of matrix multiplication in GPUs, GreenMM \cite{zamani2019greenmm} proposed GPU undervolting while keeping the frequency constant. Undervolting beyond minimum operational voltage leads to increase in number of faults. To address this challenge, GreenMM proposes an algorithm called Algorithm Based Fault Tolerance (ABFT). Tang et al. \cite{tang2019impact} studied the impact of GPU DVFS on performance and energy consumption of several DNNs and convolutional algorithms. They employed three GPUs with different architectures and four DNNs to conduct their experiments. Based on the results, the impact of high frequency on improving the performance and energy inefficiency of default frequency settings are explained. None of the aforementioned works studies the impact of reduced-precision instructions on performance of GPUs.

% \subsection{Reduced-Precision Instructions in GPUs}
% Previous works have only considered the inference performance \cite{ho2017exploiting, inferenceP40GPU} when studying the impact of reduced-precision instructions of GPUs on performance of DNNs. Experiments conducted by Ho et al. \cite{ho2017exploiting} show that naive replacement of FP32 code with half precision floating point instructions (FP16) cannot improve the performance as expected. To alleviate this challenge, they provided a more sophisticated FP32 to FP16 conversion framework to achieve higher performance improvement. Xu et al. \cite{inferenceP40GPU} studied the impact of reduced-precision instructions on the performance of AlexNet and GoogleNet. Using TensorRT and P40 GPUs, they quantified the speedup and throughput improvement of INT8 over FP32. Compared to these two works, Our experiments are more extensive because we consider more metrics such as power or GOPs/J, in addition to performance. We also study the effect of DVFS on the performance of the networks as well.

\section{Conclusion} \label{sec:conclusion}
In this paper, we evaluated the energy-efficacy of CNN workloads running on off-the-shelf FPGA and GPU accelerators, using supply voltage underscaling techniques. We also evaluated the effect of frequency scaling, as well as architectural quantization techniques on the performance, power, and energy of such accelerators. We found that combining frequency scaling and reduced-precision instructions can significantly improve the energy-efficiency of GPU, while suffering from negligible accuracy loss. Moreover, comparing the GPU and FPGA, we observed that unlike FPGA that achieves best results for various metrics at highest or lowest frequency level, GPU achieves the best results for some metrics such as GOPs/J at a frequency level between lowest and highest. Therefore, it is necessary to explore the entire frequency state space of different GPUs to find the best results for a specific metric.

%\begin{acks}
%o Robert, for the bagels and explaining CMYK and color spaces.
%end{acks}

\bibliographystyle{IEEEtran}
\bibliography{references}


\end{document}